\documentclass[
english,			
french,				
spanish,			
brazil,				
]{article}

\usepackage{arxiv}
\usepackage[utf8]{inputenc} 
\usepackage[T1]{fontenc}    
\usepackage{hyperref}       
\usepackage{url}            
\usepackage{booktabs}       
\usepackage{amsfonts}       
\usepackage{nicefrac}       
\usepackage{microtype}      
\usepackage{lipsum}		
\usepackage{graphicx}
\usepackage{amssymb}
\usepackage{amstext}
\usepackage{amsmath}
\usepackage{amsthm}
\usepackage{subfig}  
\usepackage{stfloats}

\newcommand{\rvx}{\textbf{\emph{x}}}
\newcommand{\rvy}{\textbf{\emph{y}}}

\newcommand{\rvw}{\textbf{\emph{w}}}

\newcommand{\vax}{\textbf{\emph{x}}}
\newcommand{\vay}{\textbf{\emph{y}}}

\newcommand{\var}{\textbf{\emph{r}}}
\newcommand{\vaw}{\textbf{\emph{w}}}

\title{An exploratory time series analysis of total deaths per month in Brazil since 2015}

\author{ \href{https://orcid.org/0000-0003-3557-1957}{\includegraphics[scale=0.06]{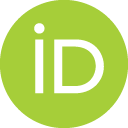}\hspace{1mm}Alexandre Barbosa de Lima} \\
	Biomedical Engineering \\
	Faculty of Sciences and Technology \\
	Pontifical Catholic University of S\~ao Paulo \\
	\texttt{ablima@pucsp.br} \\
}

\hypersetup{
pdftitle={A template for the arxiv style},
pdfsubject={q-bio.NC, q-bio.QM},
pdfauthor={David S.~Hippocampus, Elias D.~Striatum},
pdfkeywords={First keyword, Second keyword, More},
}

\begin{document}
\maketitle

\begin{abstract}
In this article, we investigate the historical series of the total number of deaths per month in Brazil since 2015 using time series analysis techniques, in order to assess whether the COVID-19 pandemic caused any change in the series' generating mechanism. The results obtained so far indicate that there was no statistical significant impact.
\end{abstract}

\keywords{COVID-19 \and Time series analysis \and Spectral analysis}

\section{Introduction}\label{sec:intro}

Brazil is a country of continental dimensions, with a territorial extension of 8,510,295.914 $\text{Km}^2$, and an estimated population of 211,959,316 people, which is distributed in a federation constituted by $26$ States and the Federal District (the country has $5,570$ cities) \cite{ibge}.
Until the time of this writing, the Brazilian Federal Ministry of Health has recorded 3,622,861 Sars-CoV-2 case reports and 115,309 deaths \cite{minsaude} caused by the COVID-19 pandemic \cite{WHO}.

According to the Coronavirus Resource Center of the Johns Hopkins University (JHU) \cite{JHU}, Brazil is the second country most affected by COVID-19 in the world, both in number of cases and in number of deaths.

This article aims to investigate the historical series of the total number of deaths using time series analysis techniques. From now on, such series will be called historical series. More specifically, we want to assess the impact of COVID-19 on the evolution of the historical series. 

The analysis was performed using the \texttt{R} software, version 4.0.2 \cite{R}. The developed \texttt{R} code, as well as the database in Excel spreadsheet format, are available for public consultation  and auditing on GitHub \cite{GitHub}.

The database used is that made available online by the Transparency Portal of the Civil Registry Offices of Brazil \cite{database}, which consolidates the amount of birth, marriage and death certificates available in Brazil. Online data has been available since January 2015.
Brazilian registries are regulated by the National Council of Justice (CNJ), which is a public institution headquartered in Brasília, Federal District, that aims to improve the work of the Brazilian judicial system, especially with regard to administrative and procedural control and transparency \cite{cnj}. The president of the Brazilian Supreme Court also presides the CNJ.

The remainder of the paper is organized as follows.  We review basic concepts of time series analysis  in Section \ref{sec:background}. Section \ref{sec:results} presents the technique used for modeling the historical series. This section also includes an exploratory data analysis and a  spectral analysis. Section \ref{sec:conclusions} summarizes the conclusions and highlights some topics for further investigation.

\section{Background}\label{sec:background}

\subsection{Time Series Basic Concepts}\label{subsec:TSBC}

\begin{figure}[ht] 
	\begin{center}
	\includegraphics[scale = 0.40]{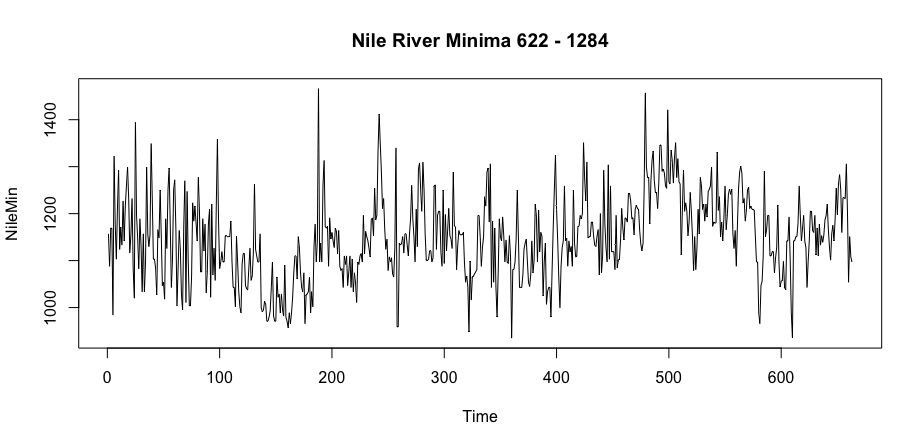} 
	\end{center}
	\caption{Nile River mimima series.}
	\label{fig:nileriverminima}
\end{figure}

In broad terms, a time series consists in a set of numbers corresponding to the observation of a certain phenomenon. Figure \ref{fig:nileriverminima} shows the Nile River mimima for the years 622 AD to 1284 AD\footnote{Yearly minimal water levels of the Nile river measured at the Roda gauge near Cairo.} \cite{omar1925}. By nature, such numbers are realizations of random variables. In general, a collection of random variables, $\{\vax_t\}$, indexed by $t$, is referred to as a stochastic process \cite{shumway2006}. In this paper, $t$ will be discrete and vary over the integers $t=0,\pm 1, \pm 2, \ldots$. 

Box and Jenkis \cite{box94} introduced the class of stationary ARMA$(p,q)$ (autoregressive moving average) models\footnote{In this work, we use the simplified notation $\vax_t$ to denote a discrete-time stochastic process $\{\vax_t\}$.}

\begin{equation}
	\label{def:arma}
	\vax_t - \mu = \sum_{j=1}^{p} \phi_j (\vax_{t-j}-\mu) + \vaw_{t} - \sum_{i=1}^{q} \theta_i \vaw_{t-j}, 
\end{equation}	

where $\mu=E\{\vax_t\}$ is the mean of $\vax_t$, $\{\phi_1, \phi_2,\ldots,\phi_p\}$ and $\{\theta_1, \theta_2,\ldots,\theta_q\}$ are parameters of the model, and $\vaw_t$ is a wide-sense stationary white noise process with zero mean and power $\sigma^2$, i. e., $\vaw_t \sim (0,\sigma^2)$. In a more compact form, we have

\begin{equation}
	\label{def2:arma}
	\phi(B) \vax'_t = \theta(B)\vaw_t, 
\end{equation}	

where $\vax'_t = \vax_t - \mu$, $B$ is the backward shift operator ($B\vax_t = \vax_{t-1}$), $\phi(B)$ is the autoregressive (AR) operator of order $p$

\begin{equation}
	\label{def:op-ar}
	\phi(B) = 1 - \phi_1 B - \phi_2 B^2 - \ldots - \phi_p B^p 
\end{equation}

and $\theta(B)$ denotes the moving average (MA) operator of order $q$

\begin{equation}
	\label{def:op-ma}
	\theta(B) = 1 - \theta_1 B - \theta_2 B^2 - \ldots - \theta_q B^q. 
\end{equation}	

In the rest of this paper, we will assume $\mu=0$ without loss of generality.

The process $\vax_t$ can be viewed as the output of a digital filter (ARMA filter) whose input is $\vaw_t$, with system function 

\begin{equation}
	\label{def:arma-filter}
	H(z) = \frac{\theta(B)}{\phi(B)},
\end{equation}

where $H(z)$ denotes the $z$-transform of the impulse response $h_t$ of the ARMA filter. An ARMA$(p,q)$ process $\vax_t$ is said to be \emph{wide-sense stationary} (or non-explosive)  if the poles of $H(z)$ in (\ref{def:arma-filter}) lie inside the complex unit circle ($|z|=1$), and it is \emph{invertible} if the zeros of $H(z)$ in (\ref{def:arma-filter}) lie inside the unit circle. 
The autocorrelation function (ACF) \footnote{We assume that the \emph{autocorrelation function} is given by  $\rho_h=\frac{\gamma_h}{\gamma_0}$, where $\gamma_h$ corresponds to the autocovariance of $\vax_t$ at lag $h$.} $\rho_h$  of an ARMA$(p,q)$ process shows exponentially decay to zero, i. e., at lag $h$ converges rapidly to zero as $h\rightarrow\infty$ (short memory property) \cite{box94}. 

A random process $\rvx_t$ is wide sense (or weakly) stationary if its mean is constant \cite[p. 298]{papoulis91}

\begin{equation}
	\label{eq:esa-media}
	E[\rvx_t] = \mu_{\rvx},
	\end{equation}

and if its ACF depends only on the lag $\tau=t_2-t_1$:

\begin{equation}
	\label{eq:esa-autocor}
	\rho_{\rvx}(t_1,t_2) = \rho_{\rvx}(t_1,t_1 + \tau) = \rho_{\rvx}(\tau).
\end{equation}

In the literature, it is common to use the terms time series and stochastic process interchangeably \cite{shumway2006}, \cite{zivot03}, \cite{tsay05}, \cite{lima2013}. From now on, we will only use the term \textit{time series}. The context will indicate to the reader whether it is a process or a realization of a process.

\subsection{Time Series Modeling}\label{subsec:TSM}

The modeling of a time series $\vax_t$ consists on estimating an invertible function $h(.)$, called \textit{model} of $\vax_t$, such that

\begin{equation} 
	\label{eq:mod-geral}
	\vax_t = h(\ldots,\rvw_{t-2},\rvw_{t-1},\rvw_{t},\rvw_{t+1},\rvw_{t+2},\ldots),
\end{equation}

in which $\rvw_{t} \sim \text{Independent and Identically Distributed -- IID}$ and

\begin{equation} 
	\label{eq:mod-geral-inv}
	g(\ldots,\rvx_{t-2},\rvx_{t-1},\rvx_{t},\rvx_{t+1},\rvx_{t+2},\ldots) = \rvw_{t},
\end{equation}

in which $g(.)=h^{-1}(.)$. The process $\rvw_{t}$ is the \textit{innovation} at instant $t$ and represents the new information about the series that is obtained at instant $t$. 

In practice, the adjusted model is \textit{causal}, i. e., 

\begin{equation} 
	\label{eq:mod-geral-causal}
	\rvx_{t} = h(\rvw_{t},\rvw_{t-1},\rvw_{t-2},\ldots).
\end{equation}

\begin{figure}[!htpb]
	\centering
	\includegraphics[scale = 0.50]{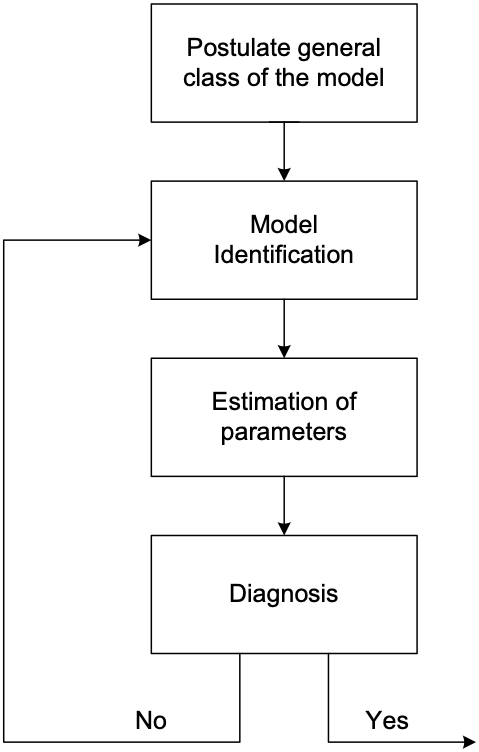}
	\caption{ Box-Jenkins' iterative cycle.}
	\label{fig:sel-modelo}
\end{figure}

The model construction methodology is based on the iterative cycle illustrated by Fig.  \ref{fig:sel-modelo} \cite{box94}:

\begin{description}
	\item[(a)] a general class of models is considered for the analysis (\textit{specification});
	\item[(b)] there is the \textit{identification} of a model, based on statistical criteria; 
	\item[(c)] it follows the \textit{estimation} phase, in which the model's parameters are obtained. In practice, it is important that the model is \textit{parsimonious}\footnote{We say that a model is parsimonious when it uses few parameters. The use of an excessive number of parameters is undesirable because the uncertainty degree of the statistical inference procedure increases with the number of parameters.};  and
	\item[(d)] at last, there is the \textit{diagnostic} of the adjusted model by means of a statistical analysis of the residual series $w_t$ (is $\rvw_t$ compatible with a white noise process?)
\end{description}

The process $\rvx_t$ of (\ref{eq:mod-geral-causal}) is \textit{linear} when it corresponds to the convolution of a process $\rvw_t \sim  \text{IID}$ and a deterministic sequence $h_t$ \cite{taqqu94}[p. 377] 

\begin{equation}
\begin{split}
\label{eq:proc-linear}
\rvx_t &= h_t \star \rvw_t = \overset{\infty}{\underset{k=0}\sum}h_k \rvw_{t-k} \\
&= \rvw_t + h_1\rvw_{t-1} + h_2\rvw_{t-2} + \ldots \\
&= (1 + h_1B + h_2B^2 + \ldots)\rvw_t \\
&= H(B)\rvw_t \\
\end{split}
\end{equation}

in which the symbol $\star$ denotes the convolution operation and $h_0=1$. 

Eq. (\ref{eq:proc-linear}) is also known as the \textit{infinite order moving average} (MA$(\infty)$) representation \cite{brockwell96}.

As the models in practice are invertible, the model of $\vax_t$  can be rewritten in an \textit{infinite order autoregressive}  (AR($\infty$)) form:

\begin{equation}
\label{def:ar-infty}
\vax_t  = \sum_{k=1}^{\infty} g_k \vax_{t-k} + \vaw_t.  
\end{equation}	

An order $p$ AR model satisfies the equation

\begin{equation}
	\label{eq:modelo-ar}
	\phi(B) \rvx_t = \rvw_t. 
\end{equation}

in which $\phi(B)$ is an order $p$ polynomial.

\subsection{Identifying AR Models}\label{subsec:IARmodels}

In practice, the order $p$ of an AR series is unknown and must be empirically specified. In this paper, we use an information criterion function \cite{tsay05} as will be explained below.

The basic idea of an ARMA model selection criterion is to choose the orders $k$ and $l$ that minimize the quantity 

\begin{equation}
	\label{eq:crit-id}
	P(k,l) = \ln{\hat{\sigma}^2_{k,l}} + (k+l)\frac{C(N)}{N},
\end{equation}

in which $\hat{\sigma}^2_{k,l}$ is a residual variance estimate obtained by adjusting an ARMA($k,l$) model to the $N$ series observations, and $C(N)$ is a function of the series size. 

The quantity $(k+l)\frac{C(N)}{N}$ is called penalty term and it increases when the number of parameters increases, while $\hat{\sigma}^2_{k,l}$ decreases.

Akaike proposed the  information criterium \cite{akaike73}, \cite{akaike74}

\begin{equation}
\label{eq:AIC}
AIC(k,l) = \ln{\hat{\sigma}^2_{k,l}} + \frac{2(k+l)}{N},
\end{equation}

known as AIC, in which $\hat{\sigma}^2_{k,l}$ is the maximum likelihood estimator of  $\sigma^2_{\rvw}$ for an ARMA($k,l$) model. 

Upper bounds $K$ and $L$ for $k$ and $l$ must be specified. Eq. (\ref{eq:AIC}) has to be evaluated for all possible $(k,l)$ combinations with $0\leq k \leq K$ $0\leq l \leq L$. In general, $K$ and $L$ are functions of $N$, for example, $K=L=\ln N$.	

For the case of AR($p$) models, (\ref{eq:AIC}) reduces to

\begin{equation}
	\label{eq:AIC-ar}
	AIC(k) = \ln{\hat{\sigma}^2_{k}} + \frac{2k}{N}, \quad k \leq K.
\end{equation}

\subsection{Estimation of AR Models}\label{subsec:EARmodels}

Having identified the AR model's order $p$, we can go to the parameters estimation phase. The methods of moments, Least Squares and Maximum Likelihood may be used \cite{percival93}, \cite{morettin04}. As, in general, the moments estimators are not good \cite{morettin04}, statistical packages as \texttt{R} and \texttt{S-PLUS} use some Least Squares or Maximum Likelihood estimator.

\subsection{ARIMA model}\label{subsec:ARIMA}

If a process which corresponds to the difference of order $d=1,2,\ldots$ of $\rvx_t$

\begin{equation}
	\label{eq:dif}
	\rvy_t= (1-B)^d \rvx_t = \Delta^d \rvx_t
\end{equation}

is stationary, then $\rvy_t$ can be represented by an ARMA($p,q$) model

\begin{equation}
	\label{eq:arma-y(t)}
	\phi(B) \rvy_t = \theta(B)\rvw_t.
\end{equation}

In this case, 

\begin{equation}
	\label{eq:arima}
	\phi(B)\Delta^d \rvx_t = \theta(B)\rvw_t
\end{equation}

is an ARIMA($p,d,q$) model and we say that $\rvx_t$ is  an ``integral'' of $\rvy_t$ \cite{morettin04} because 

\begin{equation}
	\label{eq:som}
	\rvx_t= S^d \rvy_t.
\end{equation}

The ARIMA($p,d,q$) model

\begin{equation}
	\label{eq:filtro-arima}	
	H(z)= \frac{\theta(z)}{\phi(z)(1-z^{-1})^d}
\end{equation}

is marginally stable \cite{proakis07}, as it has $d$ roots on the unit circle. Also, $\rvx_t$ of (\ref{eq:arima}) is a \textit{homogeneous non-stationary} process (meaning \textit{non-explosive}) or having  \textit{unit roots} \cite {zivot03}, \cite{tsay05}, \cite{morettin04}.

Observe that \cite{morettin04}[p.139]:

\begin{description}
	\item[(a)] $d=1$ corresponds to homogeneous non-stationary series with respect to the level (they oscillate around a mean level during a certain time and then jump to another temporary level);
	\item[(b)] $d=2$ corresponds to homogeneous non-stationary series with respect to the trend (they oscillate along a direction for a certain time and then change to another temporary direction).
\end{description}

The ARIMA model (\ref{eq:arima}) may be represented in three ways:

\begin{description}
	\item[(a)] ARMA($p+d,q$) (similar to Eq. (\ref{def:arma})) 
	\begin{equation}
	\label{eq:arima-eqdif}
	\rvx(t) = \overset{p+d}{\underset{k=1}\sum}\varphi_k \rvx(t-k) + \rvw(t) - \overset{q}{\underset{k=1}\sum}\theta_k \rvw(t-k);
	\end{equation}
	\item[(b)] AR($\infty$) (inverted format), given by (\ref{def:ar-infty}) or
	\item[(c)] MA($\infty$), according to (\ref{eq:proc-linear}).
\end{description}

\subsection{Random walk}

Consider the model $\rvy_t \sim I(1)$ 

\begin{equation}
	\label{eq:RW}
	\rvy_t = \rvy_{t-1} + \rvx_t,
\end{equation}

in which $\rvx_t$ is a stationary process. If we assume the initial condition $y_0$,  (\ref{eq:RW}) can be rewritten as an integrated sum

\begin{equation}
	\label{eq:RW-IntSum}
	\rvy_t = \rvy_0 + \sum_{j=1}^{t} \rvx_j.
\end{equation}

The integrated sum $\sum_{j=1}^{t} \rvx_j$ is called stochastic trend and it is denoted by  $TS_t$. Observe that

\begin{equation}
	\label{eq:RW2}
	TS_t = TS_{t-1} + \rvx_t,
\end{equation}

in which $TS_0=0$.

If $\rvx_t \sim \mathcal{N}(0,\sigma^2_{\rvx})$ in (\ref{eq:RW}), then $\rvy_t$ is known as \textit{random walk}. 

Including a constant in the right side of  (\ref{eq:RW}), we have a random walk with \emph{drift},

\begin{equation}
	\label{eq:RW-drift}
	\rvy_t = \theta_0 + \rvy_{t-1} + \rvx_t.
\end{equation}

Given the initial condition $y_0$, we can write

\begin{equation}
	\begin{split}
	\label{eq2:RW-drift}
	\rvy_t &= y_0 + \theta_0 t + \sum_{j=1}^{t}\rvx_{j} \\
	&= TD_t + TS_t 
\end{split}
\end{equation}

The mean, variance, autocovariance and ACF of $\rvy_t$ are given by \cite{morettin04}

\begin{align}
\mu_{t}             &= y_0 + t \theta_0 \\
\sigma^2(t)            &= t \sigma^2_{\rvx} \\
C_{k}(t)            &= (t-k)\sigma^2_{\rvx}\\
\rho_{k}(t)         &= \frac{t-k}{t}.  
\end{align}

Observe that $\rho_{k}(t)\approx 1$ when $t>>k$ and the literature states that the random walk has ``\emph{strong memory}'' \cite{tsay05}.

The random walk's Sample Autocorrelation Function (SACF) decays linearly for large  lags. 

\subsection{Spectral Analysis of Random Signals}\label{subsec:spectral}

Spectral analysis is a well-established research area \cite{percival93}. However, the estimation of the power spectrum of a signal is not a trivial matter. There are two classes of spectral analysis techniques currently in use:  parametric (or model-based) and nonparametric analysis. Both methods are used in this work.

The fundamental idea of parametric spectral analysis is fairly simple. The parametric approach assumes that the signal satisfies a generating model, such as an $\text{AR}(p)$, with known functional form and then proceed by estimating the parameters in the assumed model \cite{percival93}, \cite{stoica05}. The most widely used form of parametric Power Spectral Density (PSD) estimation uses an $\text{AR}(p)$ model \cite{percival93}.

Let us now consider the nonparametric (or classical) method. Then, the estimation of PSD of a time series $\rvx_t$ can be made using periodogram methods based on the Discrete Fourier Transform (DFT), which can be efficiently calculated by a Fast Fourier Transform (FFT) Algorithm. 

In the sequence we present the periodogram method.

Consider a time series $\rvx_t$ with $N$ values or samples, i. e., $\rvx_t=0$ outside the time interval $0 \leq t \leq N-1$. In some cases of interest, we consider that $\rvx_t$ has a size $N$, even if its actual size is $M \leq N$ (in such cases the series $\rvx_t$ must be completed with $ (N-M) $ zeros (zero padding).

The equation (\ref{eq:tdf}) is the DFT of $\rvx_t$: 

\begin{equation}
\label{eq:tdf}
X[k]= \left\{
\begin{array}{ll}
\overset{N-1}{\underset{t=0}\sum} \rvx_t e^{-j 2\pi \frac{k}{N} t}, & 0\leq k\leq N-1\\
0, & \text{otherwise}
\end{array}
\right.
\end{equation}

The PSD of $\rvx_t$ can be estimated by calculating the periodogram, given by

\begin{equation}
	\label{eq:pgram}
	P_x(f_k) = \frac{|X[k]|^2}{N}
\end{equation}

where $X[k]$ denotes the DFT of $\rvx_t$, and 

\[  f_k=0,\left(\frac{1}{N}\right), \ldots, \left(\frac{k}{N}\right), \ldots, \left(\frac{N-1}{N}\right)
\]

The periodogram is an asymptotically unbiased spectral estimator of the PSD of a random signal \cite{percival93}, \cite{stoica05}. Its main problem lies in its large variance. In other words, the periodogram is inconsistent (i. e., the dispersion of the estimates is independent of $N$). This motivates the use of a ``refined periodogram method'', like the Daniell method.

It is possible to show that the periodogram values $P_x(f_k)$ are asymptotically uncorrelated random variables \cite{percival93}. Thus we may reduce its large variance by weight averaging the periodogram over small intervals centered on the current frequency $f_k$. The practical form of the Daniell estimate can be performed using the FFT. This work uses a Daniell kernel for nonparametric PSD estimation. For further details, please refer to \cite{percival93} and \cite{stoica05}.

\section{Experimental Results}\label{sec:results}

\subsection{Time Series Regression Modeling}\label{subsec:TSRM}

\begin{figure}[ht] 
	\begin{center}
		\includegraphics[scale = 0.40]{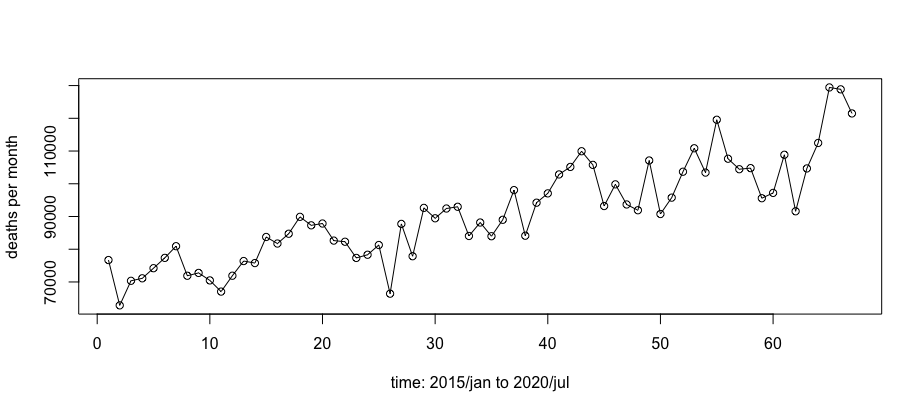} 
	\end{center}
	\caption{Historical series in Brazil from January 2015 to July 2020.}
	\label{fig:serie-original}
\end{figure}

Figure \ref{fig:serie-original} shows the historical series in Brazil from January 2015 to July 2020 ($67$ samples).

The series in the Fig. \ref{fig:serie-original} strongly suggests that we can specify a model of the type \cite{shumway2006}[p. 58]

\begin{equation}
	\label{eq:linear-trend-model}
	\vax_t  = \mu_t + \vay_t
\end{equation}

where $\vax_t$ are the observations, $\vay_t$ is a stationary process, and $\mu_t $ denotes a linear trend given by the regression model

\begin{equation}
	\label{eq:linearmodel}
	\mu_t = \beta_0 + \beta_1 t
\end{equation}

in which $\beta_0$ and $\beta_1$ are the intercept and the slope parameters.

\begin{figure}[ht] 
	\begin{center}
		\includegraphics[scale = 0.40]{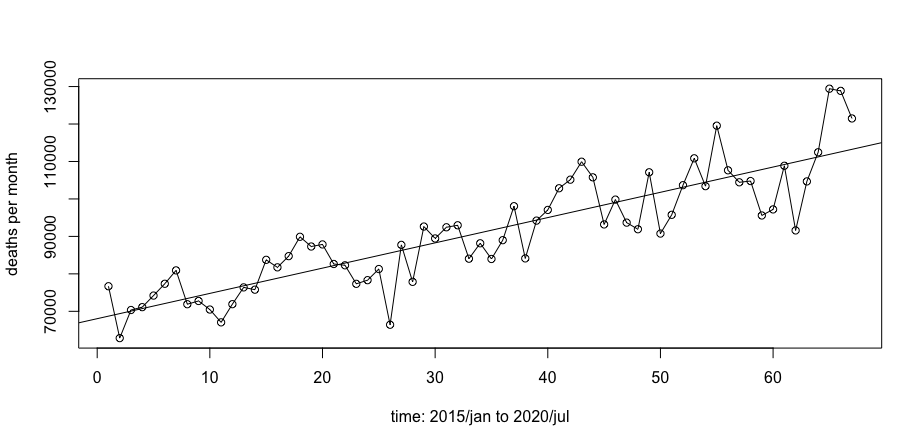} 
	\end{center}
	\caption{Time series with the superimposed linear regression model.}
	\label{fig:serie-original-fit}
\end{figure}

Table \ref{tab:estimated-model-mu} shows the estimated coeffcientes for (\ref{eq:linearmodel}) and its $p$-values, which are negligible. 
The goodness of fit is summarized by the $R^2$ of the regression \cite{zivot03}[p. 169]. Note that the  $R^2$ statistics given by Table \ref{tab:regression} indicate that the proposed model for $\mu_t$ explain approximately $75.8\%$ of the total variability of the data. The great value of the F-statistic and the neglibible model $p$-value show that we can not reject the null hypothesis of linear regression. Thus, we can consider a linear model for (\ref{eq:linearmodel}) to be statistically significant given the statistical significance level of 0.01.  Figure \ref{fig:serie-original-fit} shows the time series with the superimposed linear regression model.

\begin{table}[htp]
	\caption{Estimated model for $\mu_t$.}
	\centering
	\begin{tabular}{c c c c c c c c c} \hline\hline  
		$\widehat{\beta_0}$		 & $p$-value of  $\widehat{\beta_0}$  &  $\widehat{\beta_1}$       & $p$-value of  $\widehat{\beta_1}$ \\ \hline
		$68,036.6$                        &   $< 2.2e^{-16}$                                          &  674.7                                   &  $< 2.2e^{-16}$				 \\ \hline\hline
	\end{tabular}
	\label{tab:estimated-model-mu}
\end{table}

\begin{table}[htp]
	\caption{Adequacy of the regression model for $\mu_t$.}
	\centering
	\begin{tabular}{c c c c c c c c c} \hline\hline  
		R-squared		 & F-statistic 	    &  model  $p$-value     		\\ \hline
		$0.7575$         & $203.1$          & $< 2.2e^{-16}$      \\ \hline\hline
	\end{tabular}
	\label{tab:regression}
\end{table}

To verify that the model (\ref{eq:linearmodel}) is appropriate, it is also necessary to investigate the residuals. This is what is called residual analysis.

The residuals of (\ref{eq:linearmodel} ) correspond to the discrepancies between the observed values ($\mu$) and the values adjusted ($ \hat{\mu}$) by the model. The i-th residual is given by

\begin{equation}
	\label{eq:residuo}
	\hat{e}_i = \mu_i - \hat{\mu}_i.
\end{equation}

\begin{figure}
	\centering
	\subfloat[]{
		\label{fig:residuals-fitted}
		\includegraphics[scale = 0.25]{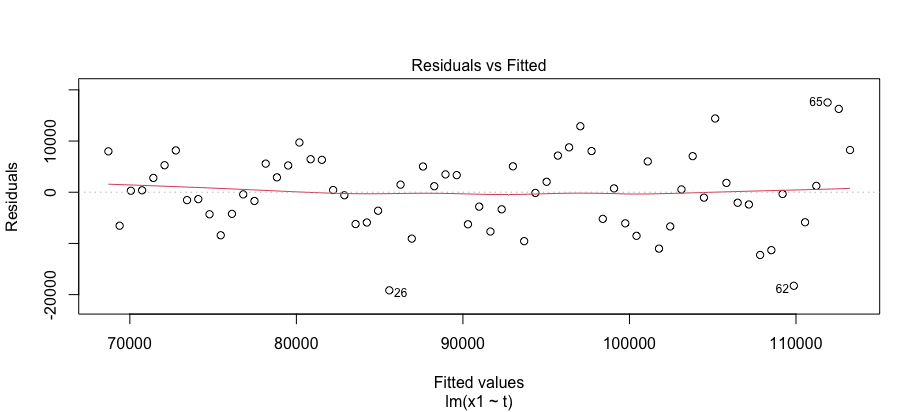}}
	\hspace{-1pt}
	\subfloat[]{
		\label{fig:qqplot-residuals}
		\includegraphics[scale = 0.25]{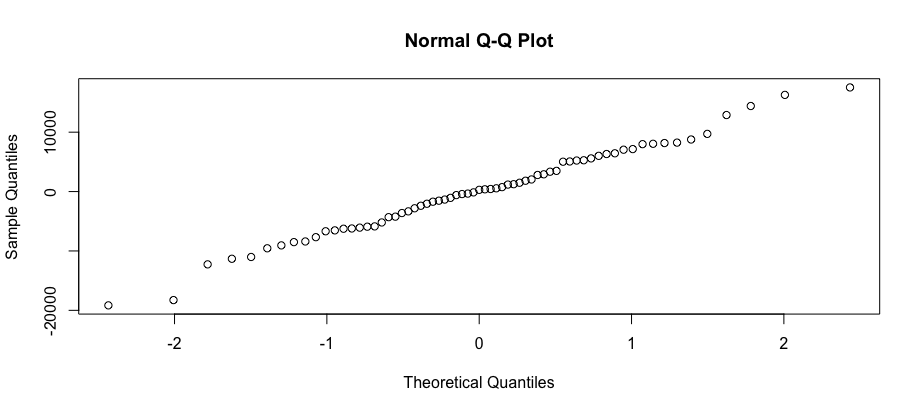}}
	\caption[]{\subref{fig:residuals-fitted} and \subref{fig:qqplot-residuals}: residuals vs fitted model and residuals QQ-plot, respectively.}
	\label{fig:resid-analysis}%
\end{figure}

Figure \ref{fig:resid-analysis} shows  the residuals vs fitted model and the Quantile-Quantile plot (QQ-plot) of the residuals, which suggests that they are normally distributed.

Table \ref{tab:resid-tests} shows a residual diagnostics using the Jarque-Bera and Shapiro-Wilks tests \cite{zivot03}[p.61] for the null hypothesis of normality of the residuals. The null of normality is not rejected using either tests.

\begin{table}[htp]
	\caption{residual diagnostics.}
	\centering
	\begin{tabular}{c c |c c} \hline\hline  
		\multicolumn{2}{c|}{Jarque-Bera test}   & \multicolumn{2}{c}{Shapiro-Wilk test}         \\ 
		      Statistic            &      $p$-value               & Statistic                  &  $p$-value                               \\ \hline
		       $0.12762$      &      $0.9382$               &  $0.9906$               &  $0.8968$                               \\\hline\hline
	\end{tabular}
	\label{tab:resid-tests}
\end{table}

At this point, we can conclude that:

\begin{itemize}
	\item there is no statistical evidence that COVID-19 affected the deterministic linear trend of the historical series, i. e., on average, the monthly growth in the number of deaths, which is approximately $675$ deaths/month, did not change since the first recorded death in Brazil on March 16, 2020; and
	\item there is no change point in the deterministic linear trend of the historical series.
\end{itemize}

\subsection{Exploratory Data Analysys}\label{subsec:EDA}

The first step in exploratory analysis is to remove the deterministic trend of Eq.(\ref{eq:linearmodel}) \cite{shumway2006}[p. 58]. There are two alternatives: remove the line estimated by the regression or take the first difference in the series. As our goal is to coerce the data to (a possible) stationarity, then differencing may be more appropriate \cite{shumway2006}[p. 61].  The first two samples of the historical series were discarded so that the series corresponding to the first difference has 64 points, that is, $2^6$ points, which facilitates the spectral analysis with FFT.

The first difference can be denote as

\begin{equation}
	\label{eq:first-diff}
	\Delta \rvx_t = \rvx_t - \rvx_{t-1} = \var_t.
\end{equation}

\begin{figure}
	\centering
	\subfloat[]{
		\label{fig:serie-detrend-demean}
		\includegraphics[scale = 0.23]{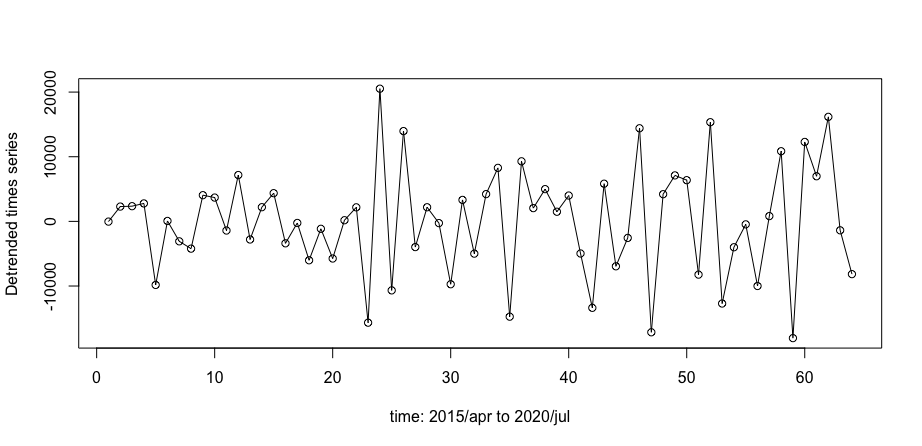}}
	\hspace{-1pt}
	\subfloat[]{
		\label{fig:SACF-serie-detrend-demean}
		\includegraphics[scale = 0.25]{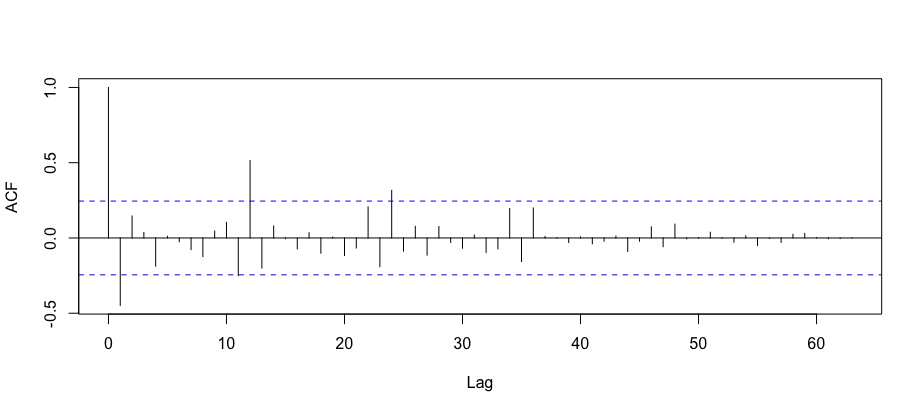}}
	\caption[]{\subref{fig:serie-detrend-demean} and \subref{fig:SACF-serie-detrend-demean}: first difference time series and its SACF, respectively.}
	\label{fig:first-diff}
\end{figure}

Figure \ref{fig:first-diff} shows the series $\var_t$ (we also demeaned the series) and its SACF. 

\begin{figure}
	\centering
	\subfloat[]{
		\label{fig:histogram}
		\includegraphics[scale = 0.25]{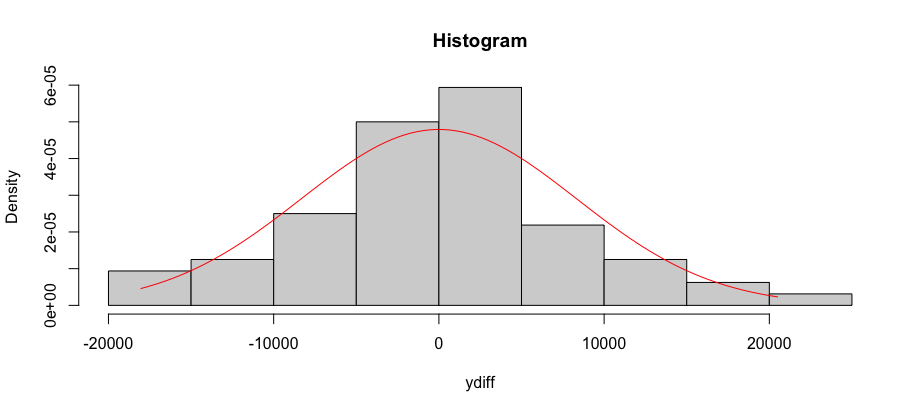}}
	\hspace{-1pt}
	\subfloat[]{
		\label{fig:qqplot}
		\includegraphics[scale = 0.25]{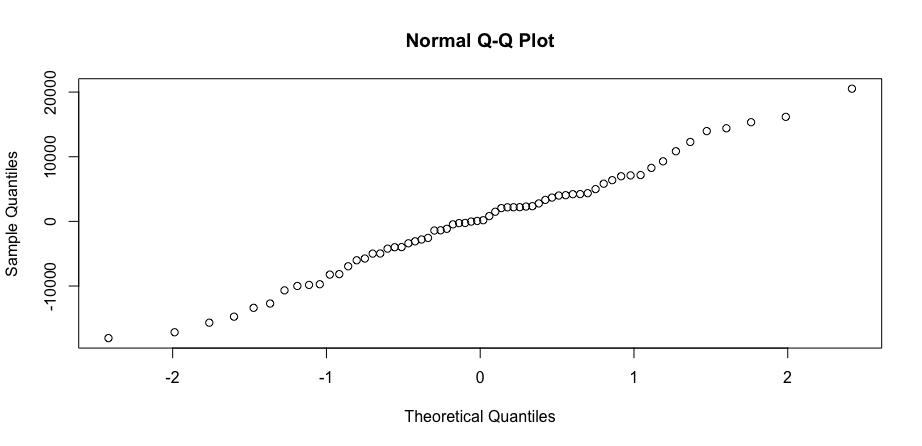}}
	\caption[]{\subref{fig:histogram} and \subref{fig:qqplot}: histogram with a superimposed normal distribution (in red) and its Q-Q plot, respectively.}
	\label{fig:first-diff-hist-qqplot}
\end{figure}

Figure \ref {fig:first-diff-hist-qqplot} shows the histogram of $\var_t$ with a superimposed normal distribution and its Q-Q plot, which suggests that $\var_t$  follows a normal distribution.

\begin{figure}
	\centering
	\subfloat[]{
		\label{fig:pgram-daniell}
		\includegraphics[scale = 0.25]{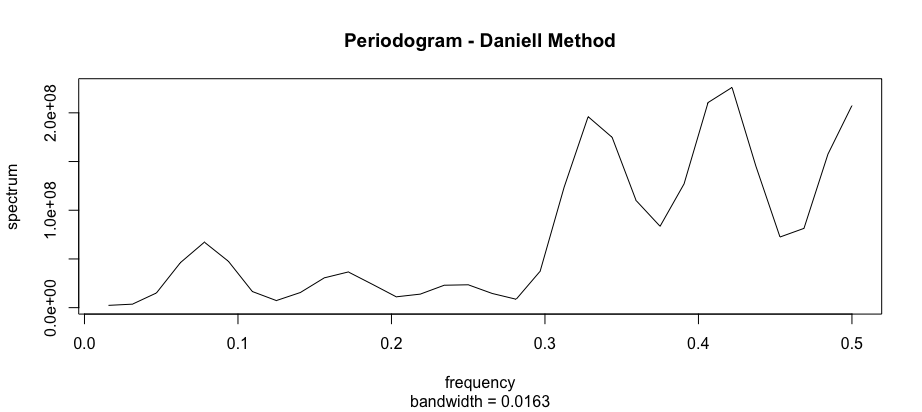}}
	\hspace{-1pt}
	\subfloat[]{
		\label{fig:psd-ar11}
		\includegraphics[scale = 0.25]{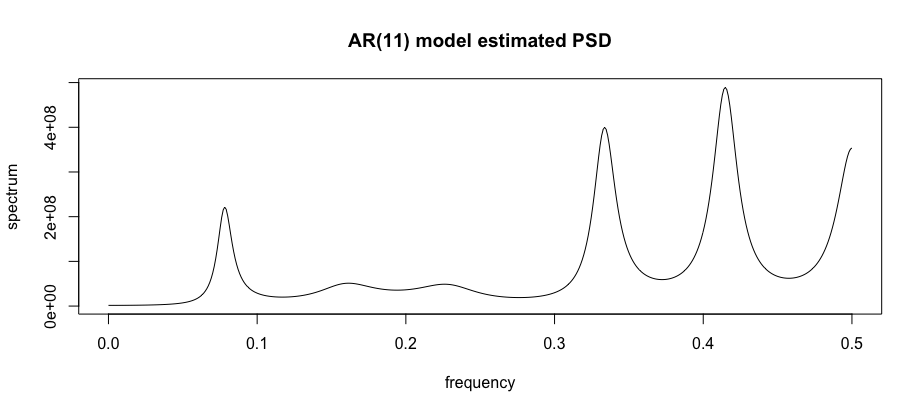}}
	\caption[]{\subref{fig:pgram-daniell} and \subref{fig:psd-ar11}: periodogram (Daniell Method) and estimated PSD for the $\text{AR}(11)$ model, respectively.}
	\label{fig:PSD}
\end{figure}

Figure \ref{fig:PSD} shows the smoothed periodogram using the Daniell method and the PSD for the estimated model of $\var_t$, which is an AR($11$).

We used the Kwiatkowski-Phillips-Schmidt-Shin (KPSS) test \cite{kpss1992} for the null hypothesis that $\var_t$ is level stationary, i. e., that the series is I($0$). We obtained a $p$-value of $0.1$, which means that one can not reject that $\var_t$ is level stationary.

Finally, but not least, an informal analysis of Fig. \ref {fig:first-diff} suggests that the variance of $\var_t$ has a change point around $t=22$ (2017/jan). However, if this really happened, which we cannot guarantee with the techniques employed in this article, ocurred long before the outbreak of COVID-19 in Brazil. This possible change point motivates a time-frequency domain analysis using wavelets, as the localized nature of wavelet coefficients allows one to analyze the evolution of the series variance over time \cite{percival93}. This will be investigated in future work.

\section{Conclusions and Future Work}\label{sec:conclusions}
In this paper, we presented an exploratory time series analysis of the historical series of the total number of deaths per month in Brazil since 2015. Our preliminary results indicate that: 

\begin{itemize}
	\item there is no statistical evidence that COVID-19 affected the deterministic linear trend of the historical series, i. e., on average, the monthly growth in the number of deaths, which is approximately $675$ deaths/month, did not change since the first recorded death in Brazil on March 16, 2020; 
	\item there is no change point in the deterministic linear trend of the historical series; 
	\item there is significant statistical evidence that the first difference time series is stationary; and
	\item there is no statistical evidence that COVID-19 provoked a change in the stochastic process that generates the time series under analysis\footnote{The random process is the series' generating mechanism.}.	
\end{itemize}

These results are thought provoking and not intuitive. COVID-19 has caused many deaths around the world. This is an indisputable fact. However, our results suggest that this disease does not have so far an additive effect on the total number of deaths per month in Brazil. What would be a plausible explanation for this strange result?

In any case, further research should be carried out to confirm the results obtained.

In future work, we will analyze the historical series using wavelet methods. 

\bibliographystyle{IEEEtran}
\bibliography{references}

\end{document}